\documentclass[fleqn,twoside]{article}
\usepackage[headings]{espcrc2}

\readRCS
$Id: espcrc2.tex,v 1.2 2004/02/24 11:22:11 spepping Exp $
\usepackage{graphicx}
\usepackage[figuresright]{rotating}

\newcommand{\AmS}{{\protect\the\textfont2
  A\kern-.1667em\lower.5ex\hbox{M}\kern-.125emS}}

\title{CLEO spectroscopy results}

\author{Jonathan L. Rosner\address{Enrico Fermi Institute and Department of
        Physics, University of Chicago \\ 
        5640 South Ellis Avenue, Chicago, IL 60637 USA}}
       
\runtitle{CLEO spectroscopy results}
\runauthor{J. Rosner}

\begin{document}

\begin{abstract}
Recent contributions of the CLEO experiment to hadron spectroscopy are
presented.
\vspace{1pc}
\end{abstract}

\maketitle

\section{INTRODUCTION}

Hadron spectroscopy plays a valuable role in particle physics.  It was crucial
in validating quantum chromodynamics (QCD) and the quark substructure of
matter.  It provides a stage for understanding nonperturbative techniques,
not only in QCD but elsewhere in physics.  Hadron spectra are crucial in
separating electroweak physics from strong interaction effects, as in charm and
beauty decays.  Quarks and leptons {\it themselves} have an intricate level and
weak coupling structure for which we have no fundamental understanding.
Sharpening spectroscopic techniques may help solve this problem.

I shall present recent spectroscopy contributions from the CLEO Collaboration
based on data at the Cornell Electron Storage Ring (CESR) utilizing CLEO's
excellent particle identification and resolution.  Separate sections will treat
charmonium, charm, beauty, and upsilons.  The CLEO detector is described in
another contribution to this Conference \cite{Rosner:dal}.

\section{CHARMONIUM}

\begin{figure}
\includegraphics[width=0.98\columnwidth]{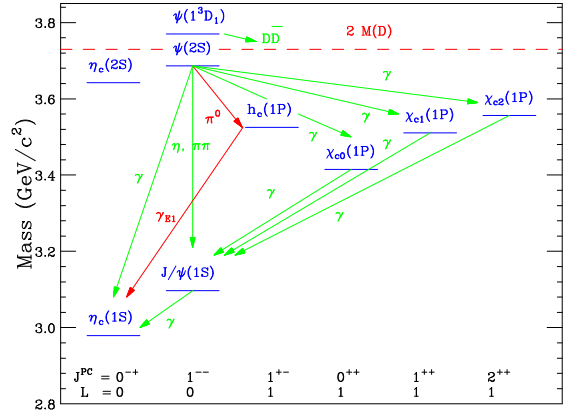}
\vskip -0.3in
\caption{The low-lying charmonium spectrum.  Bold arrows denote
$\psi(2S) \to \pi^0 h_c$, $h_c \to \gamma \eta_c$.
\vskip -0.2in
\label{fig:charmon}}
\end{figure}

The charmonium spectrum is shown in Fig.\ \ref{fig:charmon}.  Specific topics
which will be discussed are:  (1) a new measurement of ${\cal B}(J/\psi \to
\ell^+ \ell^-)$ using $\psi(2S)$ decays; (2) a study of $\psi(2S)$ decays to
baryon-antibaryon, $J/\psi X$, light hadrons, and $\pi^0 h_c$;  (3) a
remeasurement of $\Gamma(\chi_{c2} \to \gamma \gamma)$; and (4) results on
$\psi'' \equiv \psi(3770)$ decays to non-$D \bar D$ final states such as
$\pi \pi J/\psi$, $\gamma \chi_{c1}$, and light hadrons.

One can compare ${\cal B}(\psi(2S) \to \pi^+ \pi^- J/\psi \to \pi^+ \pi^-
\ell^+ \ell^-)$ with ${\cal B}(\psi(2S) \to \pi^+ \pi^- X)$ in order to
derive a value of ${\cal B}(J/\psi \to \ell^+ \ell^-)$ \cite{Li:2005ug}.
The results are ${\cal B}(J/\psi\to e^+ e^-)=(5.945\pm0.067\pm0.042)\%$,
${\cal B}(J/\psi \to \mu^+ \mu^-)=(5.960\pm0.065\pm0.050)\%$,
${\cal B}(J/\psi \to \ell^+ \ell^-)=(5.953\pm0.056\pm0.042)\%$, and
${\cal B}(e^+ e^-)/{\cal B}(\mu^+ \mu^-)=(99.7\pm1.2\pm0.6)\%$.  These values
are consistent with and more precise than current world averages
\cite{Eidelman:2004wy}.

Decays of $\psi(2S)$ to baryon-antibaryon pairs have been measured more
precisely \cite{Pedlar:2005px}.  Results are listed in Table \ref{tab:bbbar}.

\begin{table}
\caption{Branching ratios in units of $10^{-4}$ for $\psi(2S)$ decays to
baryon-antibaryon pairs \cite{Pedlar:2005px}.  The number of $\psi(2S)$ signal
events is denoted by $S$.
\label{tab:bbbar}}
\begin{center}
\begin{tabular}{l c c c} \hline \hline
Mode & $S$ & ${\cal B}(10^{-4})$ & $Q$ (\%) \\ \hline
$p \bar p$              & 557 & 2.87$\pm$0.12$\pm$0.15 & 13.6$\pm$1.1 \\
$\Lambda \bar \Lambda $ & 208 & 3.28$\pm$0.23$\pm$0.25 & 25.2$\pm$3.5 \\
$\Sigma^+ \overline{\Sigma^+}$ & 35 & 2.57$\pm$0.44$\pm$0.25 & -- \\
$\Sigma^0 \overline{\Sigma^0}$ & 58 & 2.62$\pm$0.35$\pm$0.21 & 20.7$\pm$4.2 \\
$\Xi^- \overline{\Xi^-}$ & 63 & 2.38$\pm$0.30$\pm$0.21 & 13.2$\pm$2.2 \\
$\Xi^0 \overline{\Xi^0}$ & 19 & 2.75$\pm$0.64$\pm$0.61 & -- \\
$\Xi^{*0} \overline{\Xi^{*0}}$ & 2 & 0.72$^{+1.48}_{-0.62}\pm 0.10$ & -- \\
                         & & ($< 3.2$ @90\% CL) & --\\
$\Omega^- \overline{\Omega^-}$ & 4 & 0.70$^{+0.55}_{-0.33} \pm 0.10$ & \\
                         & & ($< 1.6$ @90\% CL) & -- \\
\hline \hline
\end{tabular}
\end{center}
\vskip -0.3in
\leftline{}
\end{table}
One expects $Q \equiv {\cal B}[\psi(2S) \to f]/{\cal B}(J/\psi \to f)$
to be comparable to ${\cal B}[\psi(2S) \to \ell^+ \ell^-]/
{\cal B}(J/\psi \to \ell^+ \ell^-) =12.6 \pm 0.7\%$ (the ``12\% rule''),
since light-quark decays are presumably governed by $|\Psi(0)|^2$ as are
leptonic decays.  In fact, $Q$ is much smaller than 12\% for most VP and VT
modes, where P=pseudoscalar, V=vector, T=tensor, and severely so in some cases
\cite{Adam:2004pr,Bai:2003vf} For example, $Q(\rho \pi)$=(1.9$\pm$0.6)$\times
10^{-3}$, with a similar suppression for $K^{*\pm}K^\mp$.

The branching ratios in Table \ref{tab:bbbar} are $\sim 50\%$ higher than
current world averages \cite{Eidelman:2004wy} based on lower statistics.
Flavor SU(3) seems approximately valid for octet-baryon pair production.

New CLEO results on $\psi(2S) \to J/\psi X$ for a variety of states $X$
\cite{Adam:2005uh} are summarized in Table \ref{tab:psiX}.  The ratio of rates
for $\pi^+\pi^-/\pi^0\pi^0$ transitions is consistent with 2:1 expected from
isospin.  The $\pi^0/\eta$ ratio is $(4.1\pm0.4\pm0.1)\%$, somewhat higher than
theoretical expectations.  The CLEO branching ratios for the $\gamma \chi_{cJ}
\to \gamma \gamma J/\psi$ cascades are above those of current world averages
\cite{Eidelman:2004wy}.  They may be combined with inclusive ${\cal B}[\psi(2S)
\to \gamma \chi_{cJ}]$ branching ratios \cite{Athar:2004dn} to obtain
${\cal B}(\chi_{c0} \to \gamma J/\psi)=(2.0\pm0.2\pm0.2)\%$,
${\cal B}(\chi_{c1} \to \gamma J/\psi)=(37.9\pm0.8\pm2.1)\%$, and
${\cal B}(\chi_{c2} \to \gamma J/\psi)=(19.9\pm0.5\pm1.2)\%$.
The inclusive branching ratio for $\psi(2S) \to J/\psi X$, ${\cal B} = (59.50
\pm 0.15 \pm 1.90)\%$, is to be compared with the sum of known modes $(58.9
\pm 0.2 \pm 2.0)\%$.  Thus there is no evidence for any ``missing'' modes.
The results imply ${\cal B}(\psi' \to {\rm~light~hadrons}) = (16.9 \pm 2.6)\%$,
which, when combined with the corresponding number for $J/\psi$, leads to an
excess of $2.2 \sigma$ over ${\cal B}(\psi(2S) \to \ell^+ \ell^-)/
{\cal B}(J/\psi \to \ell^+ \ell^-)$.

CLEO has studied many exclusive multi-body final states of $\psi(2S)$
\cite{Briere:2005rc}, several of which have not been reported before.
Mode by mode, deviations
from the 12\% rule rarely amount to more than a factor of two.  The suppression
of hadronic $\psi'$ final states thus appears to be confined to
certain species such as $\rho \pi, K^* \bar K$.

\begin{table}
\caption{Branching ratios for $\psi(2S) \to J/\psi X$ \cite{Adam:2005uh}.
\label{tab:psiX}}
\begin{center}
\begin{tabular}{l c} \hline \hline
Channel & ${\cal B}$ (\%) \\ \hline
$\pi^+ \pi^- J/\psi$ & 33.54$\pm$0.14$\pm$1.10 \\
$\pi^0 \pi^0 J/\psi$ & 16.52$\pm$0.14$\pm$0.58 \\
$\eta J/\psi$        &  3.25$\pm$0.06$\pm$0.11 \\
$\pi^0 J/\psi$       &  0.13$\pm$0.01$\pm$0.01 \\
$\gamma \chi_{c0} \to \gamma \gamma J/\psi$ & 0.18$\pm$0.01$\pm$0.02 \\
$\gamma \chi_{c1} \to \gamma \gamma J/\psi$ & 3.44$\pm$0.06$\pm$0.13 \\
$\gamma \chi_{c2} \to \gamma \gamma J/\psi$ & 1.85$\pm$0.04$\pm$0.07 \\
$X J/\psi$           & 59.50$\pm$0.15$\pm$1.90 \\ \hline \hline
\end{tabular}
\end{center}
\vskip -0.3in
\leftline{}
\end{table}

The elusive $h_c(1^1P_1)$ state of charmonium has been observed by CLEO
\cite{Rosner:2005ry} via $\psi(2S) \to \pi^0 h_c \to \pi^0 \gamma \eta_c$ as
shown by the bold arrows in Fig.\ \ref{fig:charmon}.  While S-wave hyperfine
charmonium splittings are $M(J/\psi) - M(\eta_c) \simeq 115$ MeV for 1S
and $M[\psi'] - M(\eta'_c) \simeq $48 MeV for 2S levels, one expects less than
a few MeV P-wave splittings since the potential is expected to be
$\sim \delta^3(\vec{r})$ for the Coulomb-like $c \bar c$ interaction.  Lattice
QCD \cite{latt} and relativistic potential \cite{Ebert:2005jj} calculations
confirm the expectation of a small P-wave
hyperfine splitting.  One expects $M(h_c) \equiv M(1^1P_1) \simeq \langle
M(^3P_J) \rangle = 3525.36 \pm 0.06$ MeV.

Earlier $h_c$ sightings (see \cite{Rosner:2005ry} for references), based on
$\bar p p$ production in the direct channel, include a few events at $3525.4
\pm 0.8$ MeV seen in CERN ISR Experiment R704; a state at $3526.2 \pm
0.15 \pm 0.2$ MeV, decaying to $\pi^0 J/\psi$, reported by Fermilab E760 but
not confirmed by Fermilab E835; and a state at $3525.8 \pm 0.2 \pm 0.2$ MeV,
decaying to $\gamma \eta_c$ with $\eta_c \to \gamma \gamma$, reported by
E835 with about a dozen candidate events \cite{Andreotti:2005vu}.

CLEO data have been analyzed in two different ways:  exclusive, in which the
$\eta_c$ is reconstructed through one of seven decay modes, and inclusive, in
which the $\eta_c$ is not identified through its decay products.
Both analyses see a signal near $\langle M(^3P_J) \rangle$.

\begin{figure}
\includegraphics[height=0.98\columnwidth,angle=270]{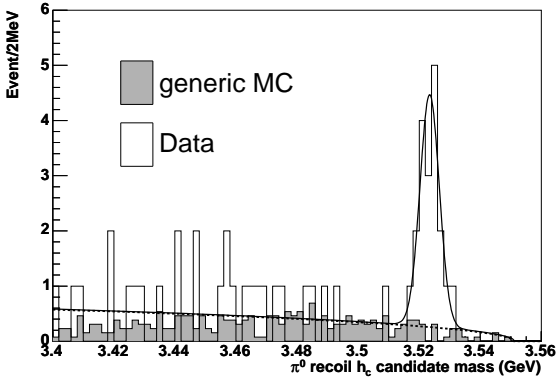}
\vskip -0.3in
\caption{Exclusive $h_c$ signal from CLEO (3 million $\psi(2S)$
decays) \cite{Rosner:2005ry}.  Data events correspond to open histogram;
Monte Carlo background estimate is denoted by shaded histogram.
The signal shape is a double Gaussian, obtained from signal Monte Carlo.
The background shape is an ARGUS function \cite{ARGUSfn}.
\vskip -0.2in
\label{fig:excl}}
\end{figure}

The exclusive signal is shown in Fig.\ \ref{fig:excl}.  A total of 19
candidates were identified, with a signal of $17.5 \pm 4.5$ events above
background.  The mass and product branching ratio for the two transitions in
Fig.\ \ref{fig:charmon} are $M(h_c) = (3523.6 \pm 0.9 \pm 0.5)$ MeV;
${\cal B}_1(\psi' \to \pi^0 h_c) {\cal B}_2(h_c \to \gamma \eta_c) =
(5.3 \pm 1.5 \pm 1.0) \times 10^{-4}$.

\begin{figure}
\includegraphics[width=0.98\columnwidth]{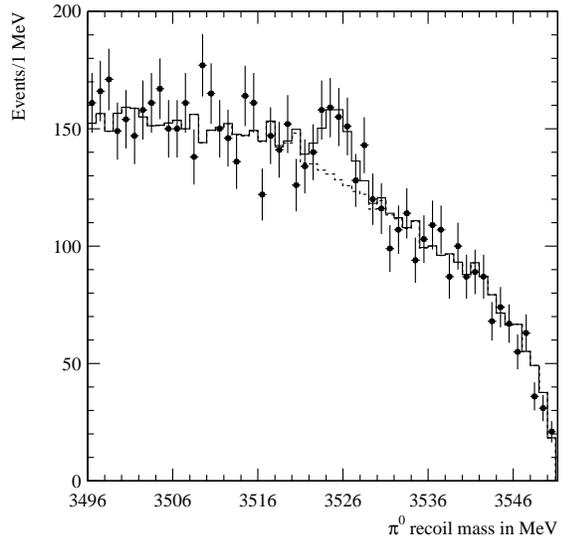}
\vskip -0.3in
\caption{Inclusive $h_c$ signal from CLEO (3 million $\psi(2S)$ decays)
\cite{Rosner:2005ry}.
The curve denotes the background function based on generic Monte Carlo plus
signal.  The dashed line shows the contribution of background alone.
\vskip -0.2in
\label{fig:incl}}
\end{figure}

The result of one of two inclusive analyses is shown in Fig.\
\ref{fig:incl}.  These yield $M(h_c) = (3524.9 \pm 0.7 \pm 0.4)$ MeV,
${\cal B}_1 {\cal B}_2 = (3.5 \pm 1.0 \pm 0.7) \times 10^{-4}$.  Combining
exclusive and inclusive results yields $M(h_c) = (3524.4 \pm 0.6 \pm 0.4)$ MeV,
${\cal B}_1 {\cal B}_2 = (4.0 \pm 0.8 \pm 0.7) \times 10^{-4}$, indicating
little P-wave hyperfine splitting in charmonium.  The $h_c$ mass is $(1.0 \pm
0.6 \pm 0.4)$ MeV below $\langle M(^3P_J) \rangle$, barely consistent with the
(nonrelativistic) bound $M(h_c) \ge \langle M(^3P_J) \rangle$
\cite{Stubbe:1991qw}.  The value of ${\cal B}_1
{\cal B}_2$ agrees with theoretical estimates of $\simeq 4 \times 10^{-4}$
obtained from $({\cal B}_1 \simeq 10^{-3}) \cdot ({\cal B}_2 \simeq 0.4)$
\cite{Kuang:2002hz,Godfrey:2002rp}.

In $\psi(2S) \to \pi^0 h_c$ one expects the $\psi(2S)$ polarization to be
transmitted to the $h_c$.  In $h_c \to \gamma \eta_c$ one then expects photons
to be distributed with respect to the beam axis according to $1 + \cos^2
\theta$.  This is confirmed by the observed signal.

CLEO has reported a new measurement of $\Gamma(\chi_{c2} \to \gamma \gamma) =
559 \pm 57 \pm 45 \pm 36$ eV based on 14.4 fb$^{-1}$ of $e^+e^-$ data at $\sqrt
{s} = 9.46$--11.30 GeV \cite{chic2}.  The result is compatible with other
measurements when they are re-evaluated using CLEO's new ${\cal B}(\chi_2 \to
\gamma J/\psi)$ and ${\cal B}(J/\psi \to \ell^+ \ell^-)$.  The errors are
statistical, systematic, and $\Delta {\cal B}(\chi_{c2} \to \gamma J/\psi$).
When the CLEO value is combined with one from Belle \cite{Abe:2002va}, the
average is $565 \pm 57$ eV.  Using the Fermilab E835 value of $\Gamma(\chi_2) =
1.94 \pm 0.13$ MeV \cite{Andreotti:2005ts} and ${\cal B}(\chi_2 \to \gamma
J/\psi) = (19.9\pm0.5\pm1.2)\%$, one finds $\Gamma(\chi_2 \to$ hadrons) =
$1.55\pm0.11$ MeV = (2.74$\pm$0.34) $\times 10^3$ $\Gamma(\chi_2 \to \gamma
\gamma)$, implying $\alpha_S(m_c) = 0.293 \pm 0.013$ if $\Gamma(\chi_2 \to$
hadrons) is dominated by the two-gluon width.  Here the QCD corrections in
\cite{Kwong:1987ak} have been used.

CLEO \cite{He:2005bs} and BES \cite{Ablikim:2004ck} have measured the $D \bar
D$ production cross sections at the peak of the $\psi''$ resonance, resulting
in values \cite{Mahlke:2005} that are somewhat less than the average
\cite{Rosner:2004wy} $\sigma(\psi'') = (7.9 \pm 0.6)$ nb of various
direct measurements.  A new value based on BES data, $\sigma[\psi'' \to ({\rm
non-} D \bar D)] = (0.72\pm0.46\pm0.62)$ nb \cite{Rong:2005it}, does not say
whether there are significant decay modes of $\psi''$ other than $D \bar D$.

Some branching ratios for $\psi'' \to X J/\psi$ are summarized in Table
\ref{tab:nonDD} \cite{Adam:2005mr}.  The value of ${\cal B}[\psi'' \to \pi^+
\pi^- J/\psi]$ found by CLEO is about 2/3 that reported by BES
\cite{Bai:2003hv}.  The entries in Table \ref{tab:nonDD} account for less than
0.5\% of the total $\psi''$ decays.

\begin{table}
\caption{Recent CLEO results on $\psi'' \to X J/\psi$ decays
\cite{Adam:2005mr}.
\label{tab:nonDD}}
\begin{center}
\begin{tabular}{|c|c|} \hline \hline
$\psi''$ mode & ${\cal B}$ (\%) \\ \hline
$\pi^+ \pi^- J/\psi$ & 0.214$\pm$0.025$\pm$0.022 \\
$\pi^0 \pi^0 J/\psi$ & 0.097$\pm$0.035$\pm$0.020 \\
    $\eta J/\psi$    & 0.083$\pm$0.049$\pm$0.021 \\
   $\pi^0 J/\psi$    & $< 0.034$ (90\% c.l.) \\ \hline \hline
\end{tabular}
\end{center}
\vskip -0.3in
\leftline{}
\end{table}

CLEO has recently reported results on $\psi'' \to \gamma \chi_{cJ}$ partial
widths, based on the exclusive process $\psi'' \to \gamma \chi_{c1,2} \to
\gamma \gamma J/\psi \to \gamma \gamma \ell^+ \ell^-$ \cite{psipprad}.  The
results are compared  in Table \ref{tab:psipprad} with some predictions
\cite{Rosner:2004wy,Eichten:2004uh}.  The exclusive analysis has no
sensitivity to $\chi_{c0}$ since ${\cal B}(\chi_{c0} \to J/\psi)$ is small.
Although the $\psi'' \to \gamma \chi_{c0}$ partial width is expected to be
high, it must be studied in the inclusive channel, which has high background,
or using exclusive hadronic $\chi_{c0}$ decays.
Even with the maximum likely $\Gamma(\psi'' \to \gamma \chi_{c0})$, one thus
expects ${\cal B}(\psi'' \to \gamma \chi_{cJ}) < {\cal O}$(2\%).

\begin{table}
\caption{CLEO results on radiative decays $\psi'' \to \gamma \chi_{cJ}$
\cite{psipprad}.
Theoretical predictions of Ref.\ \cite{Eichten:2004uh} are (a) without and
(b) with coupled-channel effects; (c) shows predictions of Ref.\
\cite{Rosner:2004wy}.
\label{tab:psipprad}}
\begin{center}
\begin{tabular}{|c|c|c|c|c|} \hline \hline
Mode & \multicolumn{3}{c|}{Predicted (keV)} & CLEO (keV) \\ \cline{2-4}
     & (a) & (b) & (c) & preliminary \\ \hline
$\gamma \chi_{c2}$ & 3.2 & 3.9 & 24$\pm$4 & $<40$ (90\% c.l.) \\
$\gamma \chi_{c1}$ & 183 & 59 & $73\pm9$ & $75\pm14\pm13$ \\
$\gamma \chi_{c0}$ & 254 & 225 & 523$\pm$12 & $<1100$ (90\% c.l.) \\ \hline
\end{tabular}
\end{center}
\vskip -0.3in
\leftline{}
\end{table}

Several CLEO analyses search for $\psi'' \to ({\rm light~
hadrons})$.  The value of $\sigma(\psi'' \to {\rm hadrons})$ also is being
re-checked.  Two analyses \cite{Adams:2005,Huang:2005} find no evidence for any
light-hadron $\psi''$ mode above expectations from continuum production except
$\phi \eta$.  Upper limits on the sum of 26 modes imply a bound ${\cal B}
[\psi'' \to$ (light hadrons)] $\le 1.8\%$.  The cross sections at 3.77 GeV
for the most part are consistent with continuum at 3.67 GeV, Both CLEO
\cite{Adams:2005} and BES \cite{LP123}, in searching for enhanced light-hadron
modes, find the $\rho \pi$ mode, suppressed in $\psi(2S)$ decays, also
suppressed in $\psi''$ decays with respect to continuum
expectations, perhaps as a result of interference between a $\psi'' \to \rho
\pi$ amplitude and continuum \cite{LP123}.

One thus can ascribe no more than a few percent of the total $\psi''$ width
to the non-$D \bar D$ decays studied thus far, including $<0.5\%$
for $J/\psi$ + (hadrons), $<0.5\%$ for $\gamma \chi_{c1,2}$, probably $<2\%$
for $\gamma \chi_{c0}$, and at most a couple of percent for light hadrons.
The question of significant non-$D \bar D$ modes of $\psi''$ remains open.
One is trying to understand a possible discrepancy of 1--2 nb in $\sigma(e^+
e^- \to \psi'')$, or ${\cal B} (\psi'') = 10$--20\%.  Could this call for more
careful treatment of radiative corrections?  A remeasurement of
$\sigma(\psi'')$ by CLEO, preferably through an energy scan, is crucial.

\section{CHARM}

CLEO has recently remeasured mass differences and widths of the singly-charmed
baryon $\Sigma^*_c(2516, J^P = 3/2^+)$ \cite{Athar:2004ni}.  Splittings
between the doubly-charged and neutral masses are quite small, in accord with
theoretical expectations.  The prediction of heavy quark symmetry that
$\Gamma(\Sigma_c^{*++})/ \Gamma(\Sigma_c^{++}) = \Gamma(\Sigma_c^{*0})/
\Gamma(\Sigma_c^{0}) = 7.5 \pm 0.1$ is borne out by the data, in which
$\Gamma(\Sigma_c^{*++})/\Gamma(\Sigma_c^{++}) = 6.5 \pm 1.3$,
$\Gamma(\Sigma_c^{*++})/\Gamma(\Sigma_c^{++}) = 7.5 \pm 1.7$.

\section{BEAUTY}

A search was performed for an energy at
which $\Lambda_b \overline{\Lambda}_b$ production might be enhanced
\cite{Besson:2004qj}.  No such enhancement was found. Upper bounds include
$R(\Lambda_b \overline{\Lambda}_b) \stackrel{<}{\sim} 0.04$ (95\% c.l.),
where $R$ refers to the cross section normalized by $\mu^+ \mu^-$ production.
Events with $\ge 1 \bar p$ and events with $\ge 1 \bar \Lambda$ did not show
any evidence of enhanced $\Lambda_b \overline{\Lambda}_b$ production just above
threshold.

\section{BOTTOMONIUM}

CLEO data continue to yield new results on $b \bar b$ spectroscopy.
New values of ${\cal B}[\Upsilon(1S,2S,3S) \to \mu^+ \mu^-] = (2.39 \pm 0.02
\pm 0.07, 2.03\pm0.03\pm0.08,2.39\pm0.07\pm0.10)\%$ \cite{Adams:2004xa} imply
lower values of $\Gamma_{\rm tot} (2S,3S)$, which will be important in updating
comparisons with perturbative QCD.  The study of $\Upsilon(2S,3S) \to \gamma X$
decays \cite{Artuso:2004fp} has provided new measurements of E1 transition
rates to $\chi_{bJ}(1P),~\chi'_{bJ}(2P)$ states.  Searches in these data for
the forbidden M1 transitions to spin-singlet states of the form $\Upsilon(n'S)
\to \gamma \eta_b(nS)~(n \ne n')$ have excluded many theoretical models.  The
strongest upper limit, for  $n'=3$, $n=1$, is ${\cal B} \le 4.3 \times 10^{-4}$
(90\% c.l.).  Searches for the lowest $b \bar b$ spin-singlet, the $\eta_b$,
using the sequential processes $\Upsilon(3S) \to \pi^0 h_b(1^1P_1) \to \pi^0
\gamma \eta_b(1S)$ and $\Upsilon(3S) \to \gamma \chi'_{b0} \to \gamma \eta
\eta_b(1S)$ \cite{Voloshin:2004hs} are being conducted.

The direct photon spectrum in $\Upsilon(1S,2S,3S) \to \gamma X$ decays has been
measured using CLEO data \cite{Besson:2005} and is used to extract the ratio of
radiative to purely gluonic decay widths.  The ratios $R_\gamma \equiv
{\cal B}(g g \gamma)/ {\cal B}(g g g)$ are found to be
$R_\gamma(1S) = (2.50\pm0.01\pm0.19\pm0.13)\%$,
$R_\gamma(2S) = (3.27\pm0.02\pm0.58\pm0.17)\%$,
$R_\gamma(3S) = (2.27\pm0.03\pm0.43\pm0.16)\%$.
$R_\gamma(1S)$ is consistent with an earlier CLEO value of
$(2.54\pm0.18\pm0.14)\%$, and the other two are first measurements.

\begin{figure}
\includegraphics[width=0.98\columnwidth]{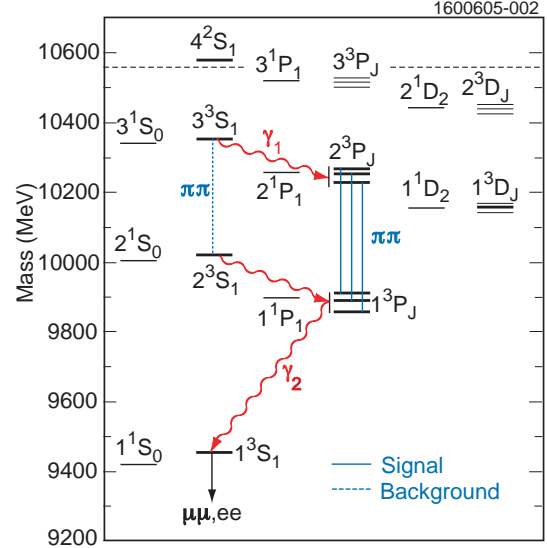}
\vskip -0.3in
\caption{Level diagram of bottomonium states illustrating the transitions
$\chi'_{bJ} \to \pi \pi \chi_{bJ}$.
\vskip -0.2in
\label{fig:chipipi}}
\end{figure}

The transitions $\chi'_b \to \chi_b \pi^+ \pi^-$ have been observed for the
first time \cite{Skwarnicki:2005pq}.  One looks for $\Upsilon(3S) \to \gamma
\to \gamma \pi^+ \pi^- \to \gamma \pi^+ \pi^- \gamma \Upsilon(1S)$ in CLEO
data, consisting of 5.8 million $\Upsilon(3S)$ events.  (See Fig.\
\ref{fig:chipipi}.)  Two methods are employed, whereby either both or only one
of the transition pions are identified, trading sample cleanliness against
statistical power.  The resulting signal event counts are 7 events above
0.6$\pm$0.2 background and 17 events above 2.2$\pm$0.6 background, repectively.
Assuming $\Gamma(\chi'_{b1} \to \pi^+ \pi^- \chi_{b1}) = \Gamma(\chi'_{b2} \to
\pi^+ \pi^- \chi_{b2})$, both are found equal to $(0.80 \pm 0.21
^{+0.23}_{-0.17})$ keV, which is in satisfactory agreement with theoretical
expectations \cite{Kuang:1981se}.  Analysis of $\chi'_b \to \pi^0 \pi^0
\chi_b$ is in progress.

\section{SUMMARY}

CLEO has contributed many recent charmonium, charm, beauty, and $b \bar b$
spectroscopy results.  The long-sought $h_c$ (the spin-singlet
P-wave ground state of charmonium) has been identified.  Its mass and
production rate confirm basic ideas about quark confinement and
isospin-violating $\pi^0$-emission transitions.  The decays of $\psi''$ are
shedding light on its nature and we look forward to much more data on this
state.  A rich CLEO program will include much further spectroscopy, such as the
study of resonances above thresholds for non-strange and strange charmed meson
pair production; $D_s$ studies; and the study of $J/\psi \to$
light-quark and glueball states.

The CLEO Collaboration gratefully acknowledges the effort of the CESR staff
in providing us with excellent luminosity and running conditions.
This work was supported by the National Science Foundation
and the United States Department of Energy.  I thank H. Mahlke-Kr\"uger for a
careful reading of the manuscript and M. Tigner
for hospitality of the Laboratory for Elementary-Particle
Physics at Cornell and the John Simon Guggenheim
Foundation for partial support.

\end{document}